\documentclass[prl,aps,twocolumn,10pt,showpacs,groupedaddress,superscriptaddress,floatfix,notitlepage]{revtex4-1}
\usepackage{amsmath,amsfonts,amssymb,graphics,graphicx,epsfig,color,times}
\usepackage[utf8x]{inputenc}
\usepackage{color}
\usepackage{bbm, dsfont} 
\usepackage{subfigure}
\usepackage{hyperref}
\usepackage{mathrsfs}
\usepackage{verbatim}
\begin{document}
\title{Disorder-assisted excitation localization in chirally coupled quantum emitters}
\author{H. H. Jen}
\email{sappyjen@gmail.com}
\affiliation{Institute of Atomic and Molecular Sciences, Academia Sinica, Taipei 10617, Taiwan}

\date{\today}
\renewcommand{\r}{\mathbf{r}}
\newcommand{\f}{\mathbf{f}}
\renewcommand{\k}{\mathbf{k}}
\def\p{\mathbf{p}}
\def\q{\mathbf{q}}
\def\bea{\begin{eqnarray}}
\def\eea{\end{eqnarray}}
\def\ba{\begin{array}}
\def\ea{\end{array}}
\def\bdm{\begin{displaymath}}
\def\edm{\end{displaymath}}
\def\red{\color{red}}
\pacs{}
\begin{abstract}
One-dimensional quantum emitters with chiral couplings can exhibit nonreciprocal decay channels, along with light-induced dipole-dipole interactions mediated via an atom-waveguide interface. When the position disorders are introduced to such atomic array, we are able to identify the dynamical phase transition from excitation delocalization to localization, with an interplay between the directionality of decay rates and the strength of light-induced dipole-dipole interactions. Deep in the localization phase, its characteristic length decreases and saturates toward a reciprocal coupling regime, leading to a system dynamics whose ergodicity is strongly broken. We also find an interaction-driven re-entrant behavior of the localization phase and a reduction of level repulsion under strong disorder. The former coincides with a drop in the exponent of power-law decaying von Neumann entropy, which gives insights to a close relation between the preservation of entanglement and nonequilibrium dynamics in open quantum systems, while the latter presents a distinct narrow distribution of gap ratios in this particular disordered system. 
\end{abstract}
\maketitle
{\it Introduction.} Localization of quantum particles in a disordered media has attracted many interests since Anderson's seminal work on the absence of spin diffusion in random lattices \cite{Anderson1958}. In his simple picture, a single quenched spin can transport between lattice sites via a spin-flip interaction, while the probability of finding this initialized spin remains finite when the random energies from the disordered lattices are introduced. Other than this single-particle localization in a noninteracting regime \cite{Wiersma1997, Schwartz2007, Lahini2008, Billy2008, Roati2008, Kondov2011, Jendrzejewski2012, Semeghini2015}, resulting from the interferences of multiple scattering paths, a broad class of closed quantum systems can present this Anderson transition (metal to insulator) \cite{Evers2008}, even under the atom-atom interactions \cite{Giamarchi1988, Fisher1989, Clement2005, Fort2005}. This further leads to recent investigations of a new dynamical phase of many-body localization \cite{Bardarson2012, Vosk2013, Yao2014, Schreiber2015, Vosk2015, Nandkishore2015, Choi2016, Smith2016, Bordia2017, Roushan2017,  Xu2018, Abanin2019, Hamazaki2019, Chiaro2019} and the nonequilibrium dynamics in interacting quantum many-body systems \cite{Gritsev2010, Polkovnikov2011}, where thermalization of both systems fails. A similar phenomenon of spatial confinement of light also emerges in multiple light scattering from many different kinds of disordered structures \cite{Wiersma2013}. For example, a free-space randomly distributed atom cloud can initiate a photon localization by strong cooperative dipole-dipole interactions \cite{Akkermans2008}. Similarly in lower-dimensional disordered photonic waveguides, a localized mode of light can be excited \cite{Topolancik2007}. 

Since a true phase of matter in quantum systems is inevitably subject to the dissipation and is often interacting with each other in either short- (hard-core bosons) or long-range distances (electrons or dipolar gases), a dynamical phase transition to localization in open interacting quantum systems is difficult to be identified and therefore is less explored. Here we focus on a one-dimensional array of two-level quantum emitters (TLQE) coupled to the photonic waveguides via the evanescent waves, which provides an alternative platform to study this universal phenomenon of localization and nonergodic dynamics. A periodic array of TLQE with the chiral couplings can exhibit the nonreciprocal decay channels \cite{Stannigel2012, Luxmoore2013, Ramos2014, Arcari2014, Mitsch2014, Pichler2015, Sollner2015, Lodahl2017, Chang2018} with the time-reversal symmetry broken \cite{Bliokh2014, Bliokh2015}, and permits an infinite-range dipole-dipole interactions in the guided modes \cite{Kien2005, Kien2008, Tudela2013, Kien2017, Solano2017}. This system offers a strong coupling regime which results in many fascinating predictions and phenomena, including entangled spin dimers \cite{Stannigel2012, Ramos2014, Pichler2015}, photon-photon correlations \cite{Mahmoodian2018}, fermionic features of subradiant states \cite{Albrecht2019, Henriet2019, Zhang2019}, subradiance dynamics \cite{Henriet2019, Jen2020_subradiance}, long-lived photon pairs \cite{Ke2019}, on-demand emission of a guided photon \cite{Corzo2019}, steady-state phase diagram \cite{Jen2020_PRR}, and photon-mediated localization \cite{Zhong2020}.

Here we introduce disorders to the periodic positions of TLQE with chiral couplings and investigate their long-time dynamics. We obtain a dynamical phase boundary from excitation delocalization to localization with an interplay between a tunable directionality of light transfer \cite{Mitsch2014} and waveguide-mediated dipole-dipole interactions. We find that the light-induced dipole-dipole interactions enable a delocalization for low disorder strengths close to the reciprocal coupling regime, similar to the interaction-facilitated thermalization of two-dimensional bosons \cite{Choi2016}. We further identify a re-entrant localization phase transition, which is also observed in many-body localization driven by the on-site interactions \cite{Schreiber2015}. This behavior coincides with a slow power-law decaying von Neumann entropy, which serves a better indicator of localized excitation than the nonequilibrating localization lengths or participation ratios across the phase boundary. Level statistics is also analyzed, where the localized phase presents a reduction of gap ratio and an increase in its fluctuations. The controllable decay rates in an atom-waveguide system can be used to differentiate the localization phase in the excited states from the classical glassy dynamics \cite{Schreiber2015, Bordia2016, Fischer2016, Levi2016, Medvedyeva2016, Luschen2017}, and our study in such strongly interacting quantum interface can shed new light in the preservation of entanglement and nonequilibrium dynamics in open quantum systems. 

\begin{figure*}[t]
\centering
  \begin{tabular}{ccc}

    \includegraphics[width=8.5cm,height=4.5cm]{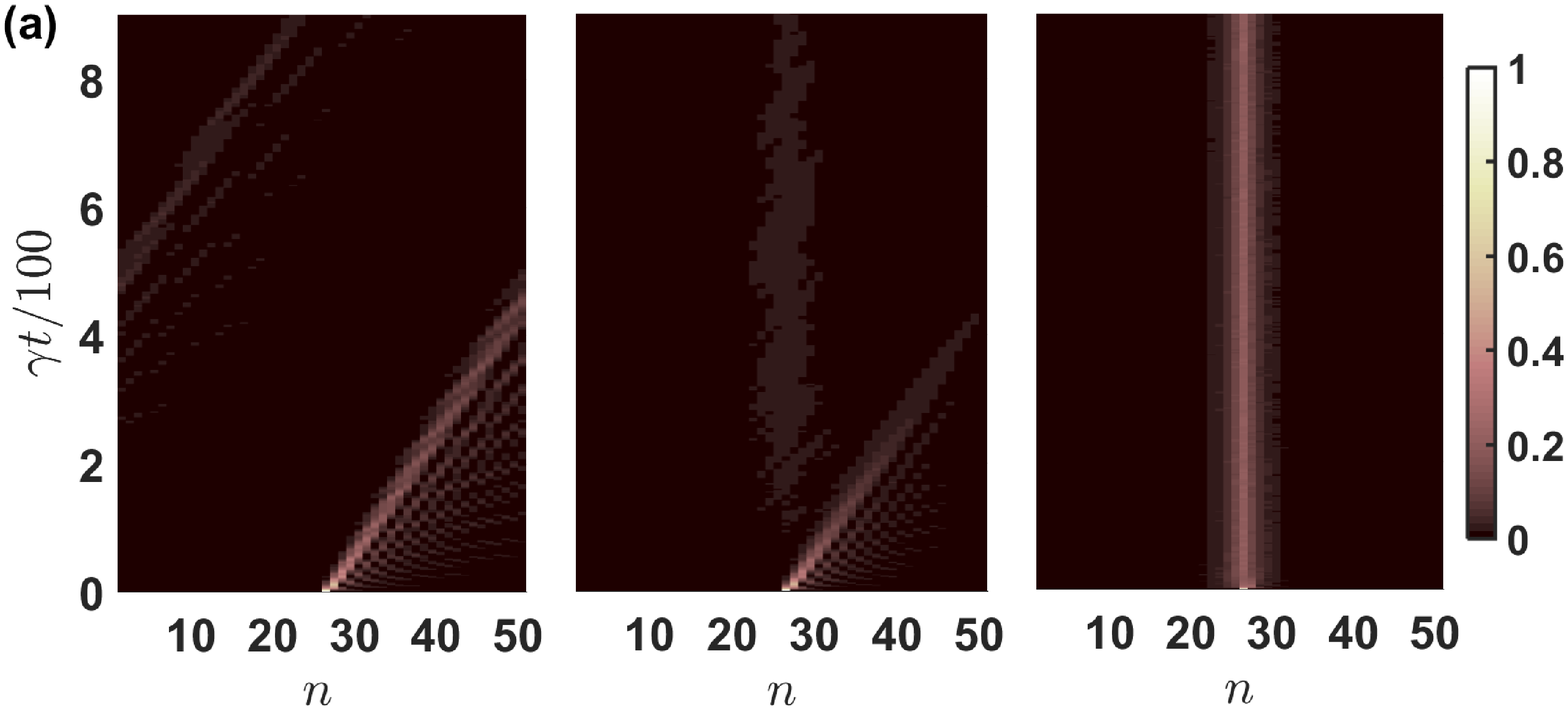}&

    \includegraphics[width=8.5cm,height=4.5cm]{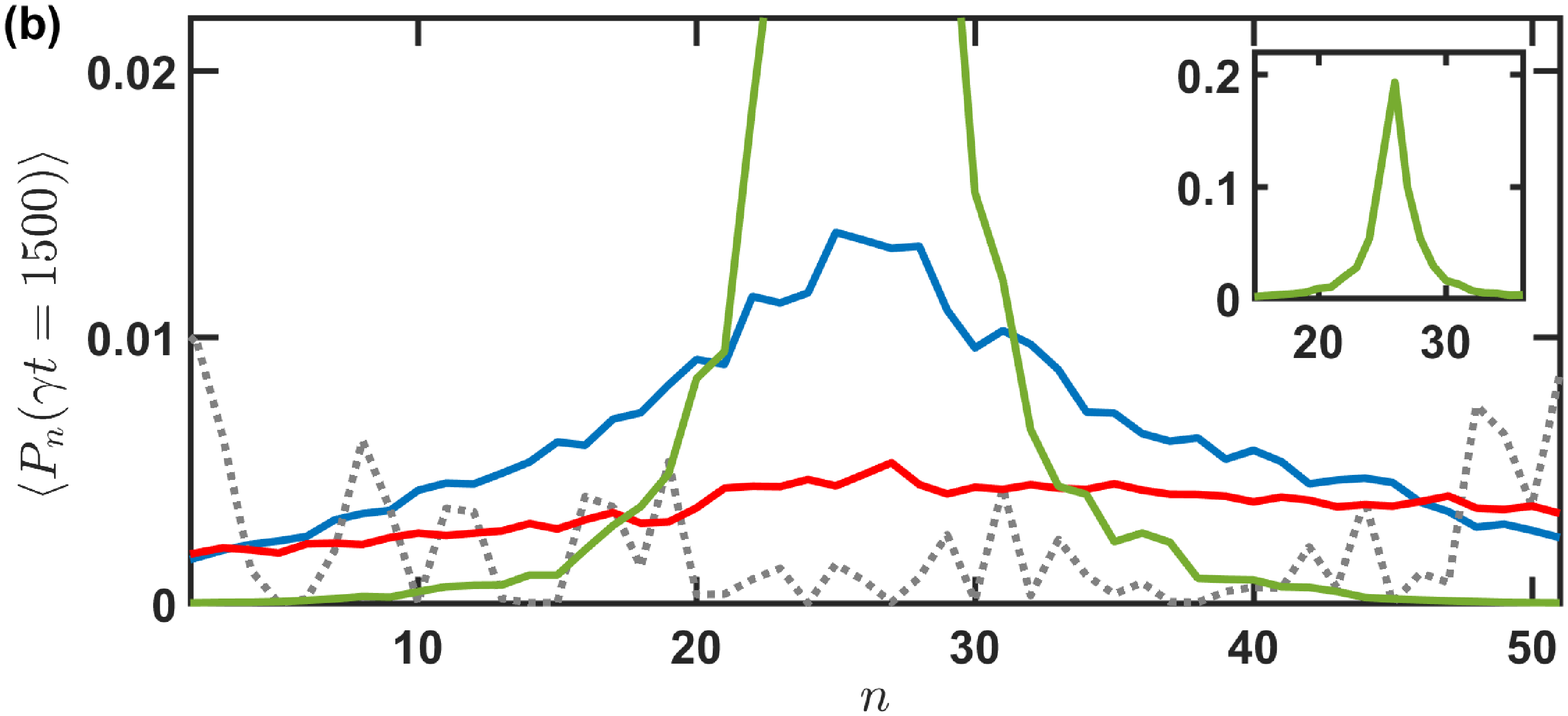}

  \end{tabular}
	\caption{Spatial and time evolutions of $\langle P_n(t)\rangle$ for $N$ $=$ $51$, $D$ $=$ $0.2$ and $\xi$ $=$ $0$ or $\pi$, and a cut at $\gamma t$ $=$ $1500$. (a) The system dynamics is shown with disorder strengths of $\bar w$ $=$ $0$, $0.02$ (near the phase boundary), and $0.2$ (deep in localization phase) in the left, middle, and right panels respectively. (b) At $\gamma t$ $=$ $1500$, an exponentially localized $\langle P_n(t)\rangle$ emerges as $\bar w$ increases from $0.01$ (red solid line), $0.02$ (blue solid line), to $0.2$ (green solid line), with a reference to the case without disorder (gray dotted line). The inset zooms out the case of $\bar w$ $=$ $0.2$ for a full picture.}\label{fig1}
\end{figure*}
 
{\it Model.} We consider a generic model in Lindblad forms to study the long-time dynamics of a one-dimensional periodic array of TLQE with chiral couplings \cite{Pichler2015}. The density matrix $\rho$ of $N$ atoms ($|g\rangle$ and $|e\rangle$ for the ground and excited states) evolves as ($\hbar$ $=$ $1$) 
\bea
\frac{d \rho}{dt}=-i[H_L+H_R,\rho]+\mathcal{L}_L[ \rho]+\mathcal{L}_R[\rho],\label{rho}
\eea
where the coherent and dissipative system dynamics are respectively determined by 
\bea
H_{L(R)} =&& -i\frac{\gamma_{L(R)}}{2} \sum_{\mu<(>)\nu}^N\left(e^{ik_s|r_\mu-r_\nu|} \sigma_\mu^\dag\sigma_\nu-\textrm{H.c.}\right)
\eea
and
\bea
\mathcal{L}_{L(R)}[\rho]=&&-\frac{\gamma_{L(R)}}{2} \sum_{\mu,\nu}^N e^{\mp ik_s(r_\mu-r_\nu)} \left(\sigma_\mu^\dag \sigma_\nu \rho + \rho \sigma_\mu^\dag\sigma_\nu \right.\nonumber\\
&&\left.-2\sigma_\nu \rho\sigma_\mu^\dag\right).
\eea
The dipole operators are $\sigma_\mu^\dag$ $\equiv$ $|e\rangle_\mu\langle g|$ with $\sigma_\mu$ $=$ $(\sigma_\mu^\dag)^\dag$, $k_s$ denotes the wave vector in the guided mode, and $\gamma_{L(R)}$ quantifies the coupling rate to the left (right) of every quantum emitter. Equation (\ref{rho}) is obtained with Born-Markov approximation \cite{Lehmberg1970} under one-dimensional reservoirs \cite{Tudela2013}, which can be treated as spin-exchange processes \cite{Dicke1954} with nonreciprocal and infinite-range dipole-dipole interactions. 

A useful factor of directionality $D$ $\equiv$ $(\gamma_R-\gamma_L)/\gamma$ \cite{Mitsch2014} defines the amount of light transfer with a normalized decay rate $\gamma$ $=$ $\gamma_R$ $+$ $\gamma_L$. $D$ $=$ $\pm 1$ presents the cascaded scheme \cite{Stannigel2012, Gardiner1993, Carmichael1993} with a unidirectional coupling, whereas a reciprocal coupling regime is reached at $D$ $=$ $0$. For an array of quantum emitters with equal interatomic distances, we use $\xi$ $\equiv$ $k_s |r_{\mu+1}-r_{\mu}|$ to quantify the strength of the light-induced dipole-dipole interactions which mediate the whole array. We add another crucial parameter of the onsite phase disorders $W_\mu$ $\in$ $\pi[-\bar w,\bar w]$ with $\bar w$ $=$ $[0,1]$, which can be established in the position fluctuations, leading to a deviation of $\xi$ in $H_{L(R)}$ and $\mathcal{L}_{L(R)}[\rho]$, or equivalently by adding the onsite disordered potentials $W_\mu|e\rangle_\mu\langle e|$ to $H_{L(R)}$.

{\it Phase boundary.} We initialize the system dynamics from a central atomic excitation, and the state of the system $|\Psi(t)\rangle$ $=$ $\sum_\mu a_\mu(t)|\psi_\mu\rangle$ under single excitation space $|\psi_\mu\rangle$ $=$ $|e\rangle_\mu|g\rangle^{\otimes(N-1)}$ evolves as 
\bea
\dot{a}_\mu(t)= &&-\gamma_R\sum_{\nu<\mu}e^{-i(\mu-\nu)\xi-i(W_\mu-W_\nu)}a_\nu(t)-\frac{\gamma}{2}a_\mu(t)\nonumber\\&&-\gamma_L\sum_{\nu>\mu}e^{-i(\nu-\mu)\xi-i(W_\nu-W_\mu)}a_\nu(t), \label{coupling}
\eea
where we have ordered the atomic positions as $r_1$ $<$ $r_2$ $<...<$ $r_{N-1}$ $<$ $r_N$. The system dynamics in the above shows a strong dependence on $\xi$ which competes with disorder strengths $W_\mu$ in determining distinct long-time behaviors. Throughout this paper, we present the converged results within time of interests, averaged over $200$ realizations of disorders. A convergence reflects in the total excitation population $\langle P_t\rangle=\sum_{n=1}^N \langle P_n\rangle$ with $P_n\equiv|a_n|^2$ and $\langle\cdot\rangle$ as an ensemble average, which is below $2\%$ deviation from the case under $2000$ realizations.

\begin{figure}[b]
\centering
\includegraphics[width=8.5cm,height=4.5cm]{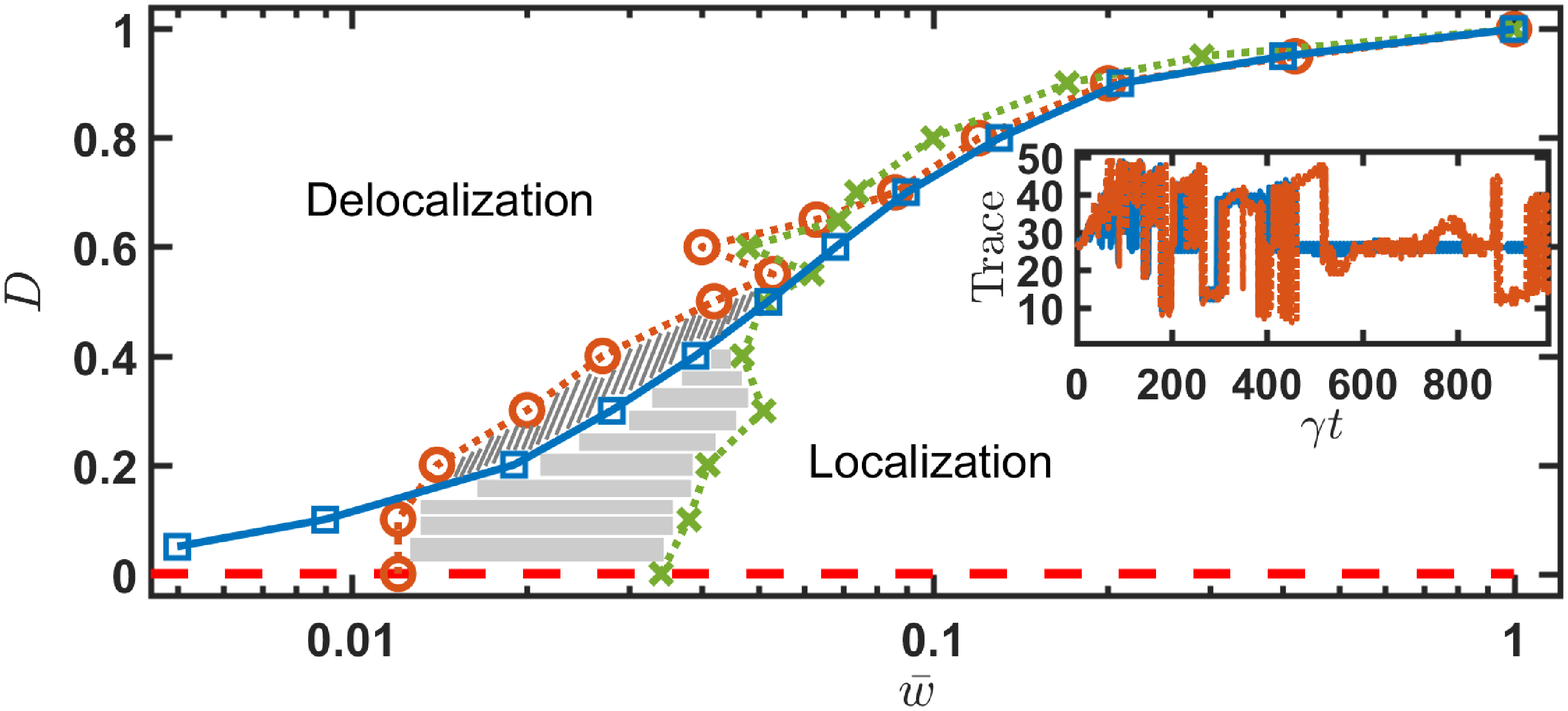}
\caption{Dynamical phase boundary of $N$ $=$ $51$ under the directionality $D$ and the disorder strengths $\bar w$ in a logarithmic scale. The phase boundary separates two distinct phases of excitation delocalization and localization for $\xi/\pi$ $=$ $0$ ($\square$), $1/8$ ($\circ$), and $1/2$ ($\times$). Both thin and thick gray-striped areas suggest a re-entrant phase transition of the localization, which can be driven by the light-induced dipole-dipole interactions. At a higher $D$, three phase boundaries start to merge and collapse to $\bar w$ $=$ $1$ as $D$ $\rightarrow$ $1$, showing that no localization is possible no matter how strong $\bar w$ is. On the other hand, at $D$ $=$ $0$ and $\xi$ $=$ $0$ or $\pi$ (red dashed line), a significant amount of this atomic excitation sustains forever due to the decoherence-free eigenmodes allowed by the system, even without disorder. We exclude this parameter regime for the converse effect of disorder. The inset shows the time traces of maximal $\langle P_n(t)\rangle$ at $D$ $=$ $0.6$ and $\xi/\pi$ $=$ $0.5$ with $\bar w$ $=$ $0.05$ (solid line) near the phase boundary in the localization side, compared to the case without disorder (dotted line).}\label{fig2}
\end{figure}

As time evolves, the central excitation without disorder in the left panel of Fig. \ref{fig1}(a) starts to move preferentially to the right since $D$ $>$ $0$. Specifically for $\xi$ $=$ $0$ or $\pi$, the main excitation populations $\langle P_n(t)\rangle$ traverse to the boundary of the lattices in a ballistic diffusion with a rate $\propto D^{-1}$, after which a repopulation appears on the other side of the lattice and propagates again. Near the phase boundary to localization, which we obtain later in Fig. \ref{fig2}, a halt of excitation transport to the end of lattices emerges along with a congregated excited-state population around the center. This represents a restoration of system's memory of the initial states and indicates a breakdown of thermalization. Further deep in the localization phase, a clear and prolonged centralized excitation can be identified. In Fig. \ref{fig1}(b), we further present their spatial distributions corresponding to the parameter regimes in Fig. \ref{fig1}(a), at a time when $P_t(\bar w=0)$ $\sim$ $0.1$. An exponentially localized excitation emerges and concentrates toward the center as disorder strength increases, in huge contrast to a thermalized $P_n(t)$ without disorder.   

It appears that a localization length (site) $n_L$ extracted from $\langle P_n(t)\rangle$ (fitted by $e^{-|n-n_c|/n_L}$ with a central atom at the $n_c$th site) can be used to estimate the dynamical phase transition to delocalization when the full-width-half-maximum of $\langle P_n(t)\rangle$ exceeds half of the lattice sites, that is $\zeta_L$ $\equiv$ $2n_L\ln(2)$ $\gtrsim$ $N/2$. However, there are two issue of using it as an estimate for phase transitions. Firstly the extracted $\zeta_L$ decreases over time owing to the interferences of spin-exchange processes through lattices, and secondly the excitation may transport to the boundary of lattices multiple times before localizing even when $\zeta_L$ $<$ $N/2$. The former makes $\zeta_L$ meaningless since near the phase boundary, $\langle P_t(\bar w\neq 0)\rangle$ approaches $P_t(\bar w= 0)$ and diminishes, while the latter violates a global transport which would forbid the localization in thermodynamics limit as $N$ $\rightarrow$ $\infty$. 

As a result, we take an operational approach to trace the transport of the maximal excited-state population of the whole lattice over time and identify the phase transition to localization when it stops traversing to the boundary it can reach without disorder. This is similar to the criterion of conductivity in electron transport, which goes to zero toward the localized phase \cite{Nandkishore2015}. We then obtain the dynamical phase boundary in Fig. \ref{fig2}, which separates the excitation delocalization and localization under the parameters of $D$ and $\bar w$. We note of a symmetry for the phase diagram as $\xi$ $\leftrightarrow$ $\pi-\xi$, which only differs in the sign of the probability amplitudes $a_n(t)$. Figure \ref{fig2} presents an asymptotic collapse of the phase boundary for various interaction strengths $\xi$ at a larger $D$, while at a lower $D$, it involves two phase areas that allow an interaction-induced re-entrance of localization phase. We will investigate this re-entrant behavior in detail later. 

At $D$ $=$ $1$, a cascaded scheme where the excitation only transfers unidirectionally, no localization is allowed no matter how strong disorder is. The system dynamics without disorder can be directly obtained as \cite{Jen2019_coherence}, 
\bea
a_{n+1}(t)=-e^{-\gamma t/2-in\xi}\int_0^t dt\sum_{n'=n_c}^{n} a_{n'}(t)e^{\gamma t/2+i(n'-1)\xi}, \nonumber\\
\eea 
where $n$ $\geq$ $n_c$ and $a_{n_c}(t)$ $=$ $e^{-\gamma t/2}$. As a consequence, $a_{n}(t)$ $\propto$ $e^{-i(n-n_c)\xi}$, where the fluctuations in $\xi$ only appear in the emitters' global phases. This can be interpreted as a lack of interference in the spin-exchange process, where the random potentials the excitation experiences in transport do not modify the dynamics of excited-state populations. For a smaller $D\lesssim 0.1$, the phase boundary moves to a higher $\bar w$ as $\xi$ increases, which indicates that a stronger disorder is required to enter the localized phase. In other words, the interactions drives the system to a more thermalized phase for low disorder strengths, which has been observed in many-body localization transition of bosons under two-dimensional disordered potentials \cite{Choi2016}. 

For a moderate $D$ $\approx$ $0.6$ with a finite $\xi$ $\neq$ $\pi$, the phase boundary shows a kink which corresponds to a collapse to localization and a revival of transport in $P_n(t)$ without disorder. We plot the time traces of maximal $\langle P_n(t)\rangle$ in the inset of Fig. \ref{fig2}, where the trace with disorder follows the one without for some time before localizing. This particular range around $D$ $\approx$ $0.6$ presents a re-organization of the localized excitation even without disorder, leading to a weaker $\bar w$ to enter the localization phase. We note that for a particular regime at $D$ $=$ $0$ with $\xi$ $=$ $0$ or $\pi$, a finite disorder destroys the highly correlated phases sustained within the decoherence-free modes and induces a decay in the excited-state population on the contrary. For a larger $N$, we find that the respective phase boundaries are pushed to a weaker $\bar w$, which enhances the effect of disorder to localization. 

\begin{figure}[t]
\centering
\includegraphics[width=8.5cm,height=4.5cm]{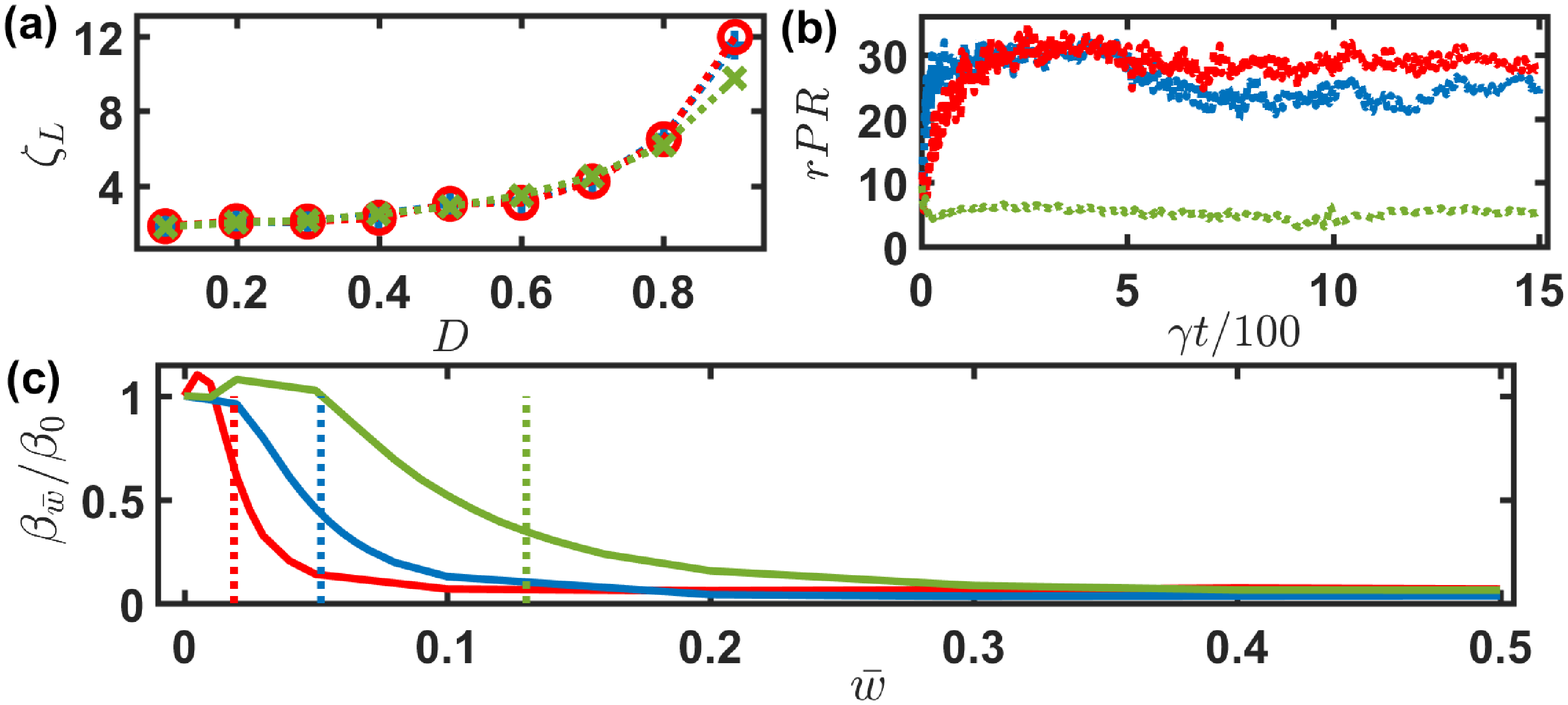}
\caption{Localization length $\zeta_L$, relative participation ratio (rPR), and exponent ratio $\beta_{\bar w}/\beta_0$ for $N$ $=$ $51$. (a) Deep in localization phase at $\bar w$ $=$ $0.5$, the average localization length fitted by an exponential function grows as $D$ increases for $\xi/\pi$ $=$ $0$ ($+$), $1/8$ ($\circ$), and $1/2$ ($\times$). (b) At $\xi/\pi$ $=$ $0$ and $D$ $=$ $0.2$, the time evolutions of rPRs distinguish each other at long time from higher values for $\bar w$ $=$ $0.01$ (red dotted line) and $0.02$ (blue dotted line), to a smaller one for $\bar w$ $=$ $0.2$ (green dotted line). (c) At $\xi/\pi$ $=$ $0$, the exponent ratios decay to their respective asymptotic values as $\bar w$ increases, and broaden as $D$ increases from $0.2$ (red solid line), $0.5$ (blue solid line), to $0.8$ (green solid line). A comparison to the phase boundary in Fig. \ref{fig2} (dotted lines) indicates a phase transition around $\beta_{\bar w}/\beta_0$ $\approx$ $0.5$ for $D$ $\lesssim$ $0.5$.}\label{fig3}
\end{figure}

{\it Localization length, participation ratio, and von Neumann entropy.} Next we study the localization length $\zeta_L$ and introduce two additional quantities that can assist the understanding of excitation localization in the system, which are participation ratios and von Neumann entropy of entanglement. Since $\zeta_L$ is not suitable to give a clear identification of the phase boundary, we investigate $\zeta_L$ deep in the localization phase instead, which gives further information of how strongly the system is localized. In Fig. \ref{fig3}(a), we obtain $\zeta_L$ by fitting it at a time when $P_t(\bar w=0)$ $\sim$ $0.1$. The $\zeta_L$ saturates as $D$ decreases, which indicates a strong nonergodic phase dominated by disorder. On the other hand as $D$ increases, $\zeta_L$ increases as well, showing a stronger dependence on the directionality of couplings over disorder strengths. 

From the participation ratio \cite{Murphy2011}, we are able to identify the extended or localized features in various parameter regimes of Fig. \ref{fig2}. We further define a property of relative participation ratio (rPR) which particularly evaluates a time-evolved measure under normalized excitation populations $\langle\tilde P_n(\bar w,t)\rangle$ at any given time $t$,  
\bea
{\rm rPR}\equiv\frac{\left(\sum_{n=1}^N \Delta \tilde P_n\right)^2}{\sum_{n=1}^N\left(\Delta \tilde P_n\right)^2}, 
\eea
where 
\bea
\Delta\tilde P_n = \big|\langle\tilde P_n(\bar w,t)\rangle - \tilde P_n(0,t)\big|\Theta\big(\langle\tilde P_n(\bar w,t)\rangle - \tilde P_n(0,t)\big), \nonumber\\
\eea
with the Heaviside step function $\Theta$ evaluating the excitation variations from a delocalized phase. In Fig. \ref{fig3}(b), we show rPRs for three different disorder strengths, across the phase boundary of $\xi$ $=$ $0$ as an example. A contrasted rPR is clearly seen deep in the localization phase, which remains a fairly low value over time, whereas near the phase boundary, the rPR in the localization side is slightly below than the one in the delocalization side, and both of them fluctuate even at longer time. The rPR serves as another informative measure on the localized excitation, but again it is not operational in distinguishing the phase boundary owing to its long-time nonequilibrating characteristic.  

The second useful quantity of von Neumann entropy gives a measure of quantum correlations throughout the whole array. We partition it at its center, and separate its left and right as $A$ and $B$. We then calculate the von Neumann entropy of entanglement as 
\bea
\langle S_{A(B)}\rangle=\langle{\rm Tr}[\rho_{A(B)}\ln\rho_{A(B)}]\rangle,\nonumber
\eea 
where $\rho_{A(B)}$ $\equiv$ ${\rm Tr}_{B(A)}[\rho]$. We particularly investigate the decaying tails of $\langle S_{A(B)}\rangle$ following the time when $S_{A(B)}$ $\sim$ $0.1$ without disorder, as a reference. We fit them by a power-law function $t^{-\beta_{\bar w}}$, where $\beta_{\bar w=0}$ characterizes how fast the entropy of entanglement falls to zero in the thermalized or ergodic phase. In Fig. \ref{fig3}(c), we show a ratio of $\beta_{\bar w}/\beta_0$ across the phase boundary for $\xi$ $=$ $0$ as an example again. For a smaller $D$, this ratio presents a sharper transition to its asymptotics $\beta_{\bar w}/\beta_0$ $<$ $0.1$ as $\bar w$ $\rightarrow$ $1$, in contrast to the broadened case for a larger $D$, and all saturate deep in the localization phase when $\bar w$ $\gtrsim$ $0.3$. We further compare them to the phase boundary of Fig. \ref{fig2} and find that $\beta_{\bar w}/\beta_0$ can be used as a measure for the onset of localization phase when $\beta_{\bar w}/\beta_0$ $\sim$ $0.5$, particularly for a low $D$. Therefore, we use this measure below to investigate further the re-entrant behavior of localization phase in Fig. \ref{fig2}. 

\begin{figure}[b]
\centering
\includegraphics[width=8.5cm,height=4.5cm]{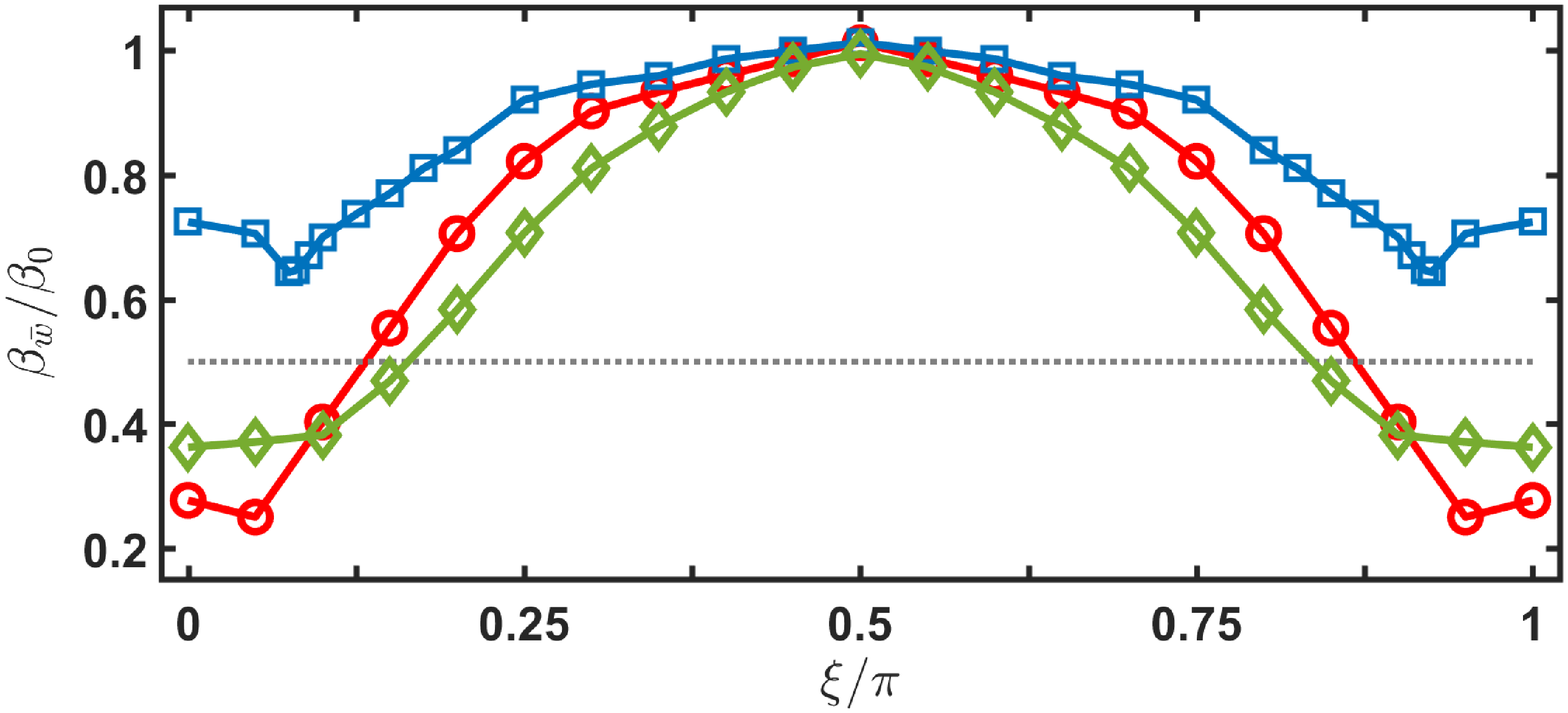}
\caption{Interaction-driven re-entrance of localization phases in the striped areas of Fig. \ref{fig2}. The re-entrance of localizations can be seen for $D$ $=$ $0.2$, $\bar w$ $=$ $0.03$ ($\circ$) and $D$ $=$ $0.3$, $\bar w$ $=$ $0.04$ ($\diamond$), in the thick striped area of Fig. \ref{fig2}, when the exponent ratios cross the referenced line at $\beta_{\bar w}/\beta_0$ $=$ $0.5$ (gray dotted line). For $D$ $=$ $0.3$, $\bar w$ $=$ $0.025$ ($\square$), in the thin striped area, a small dip shows up, which corresponds to the phase transition in Fig. \ref{fig2}, even though its ratio exceeds $0.5$.}\label{fig4}
\end{figure}

{\it Re-entrant behavior and level repulsion.} The re-entrant behavior of localization phase presents a controllable dynamical phase transition driven by interactions \cite{Schreiber2015}. In Fig. \ref{fig4}, we mark a referenced line of $\beta_{\bar w}/\beta_0$ $=$ $0.5$ as an estimate for phase transitions. We present two cases in the thick gray-striped area of Fig. \ref{fig2}, where a clear transition from the localized ($\beta_{\bar w}/\beta_0$ $<$ $0.5$) to delocalized phase ($\beta_{\bar w}/\beta_0$ $>$ $0.5$) can be seen, and the localized phase can be re-entered as $\xi$ increases. For the case in the thin gray-striped area of Fig. \ref{fig2}, we identify a dip in the ratio near the phase boundary. A small dip, instead of obviously crossing $\beta_{\bar w}/\beta_0$ $=$ $0.5$, can be attributed to a smaller phase area in Fig. \ref{fig2}.  

\begin{figure}[t]
\centering
\includegraphics[width=8.5cm,height=4.5cm]{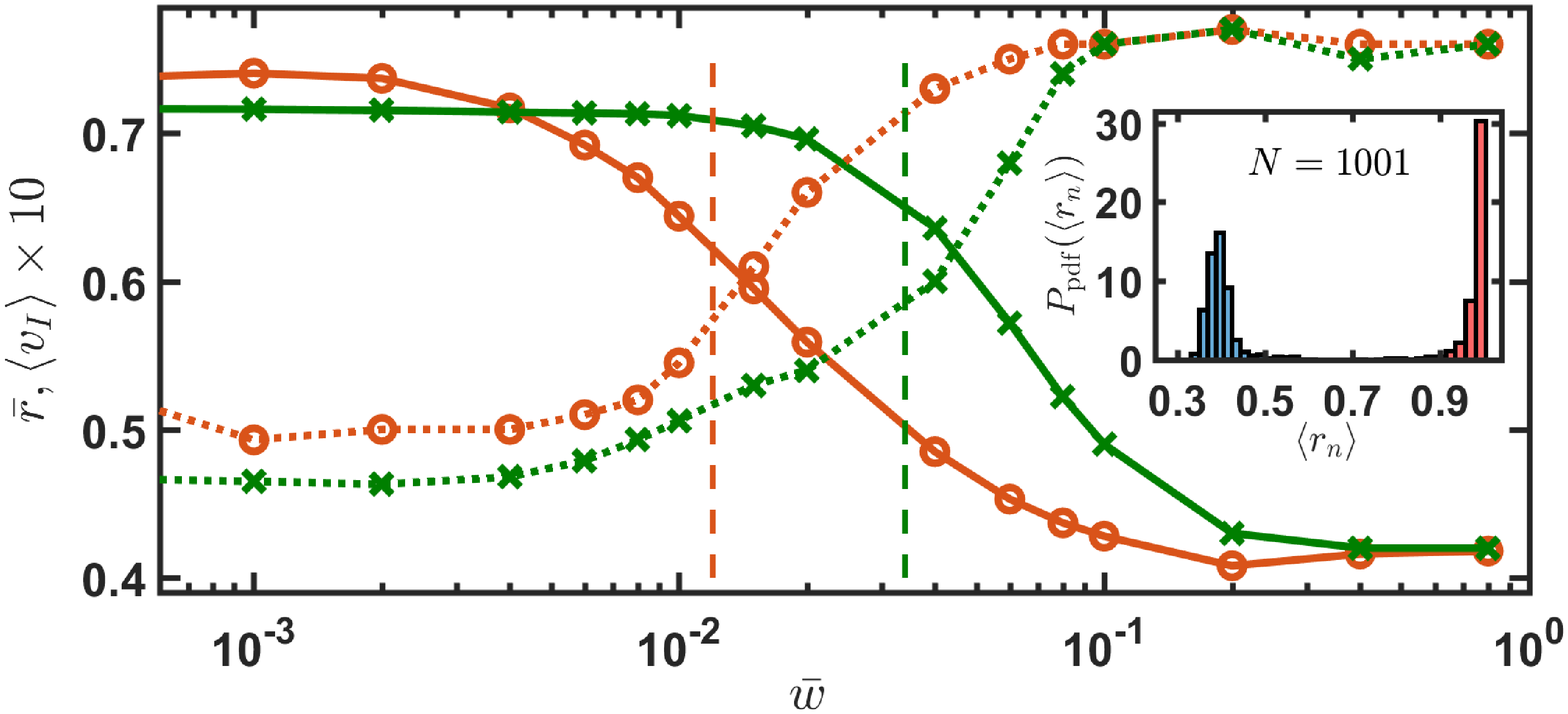}
\caption{Mean gap ratio $\bar r$ and intrasample variance $\langle v_I\rangle$ at $D$ $=$ $0$ with $N$ $=$ $51$. The gap ratios $\bar r$ for $\xi/\pi$ $=$ $1/8$ (solid line with $\circ$) and $1/2$ (solid line with $\times$) decrease as $\bar w$ increases, along with increasing fluctuations in $r_n$ (respectively with dotted lines). A comparison of phase transition to localization in Fig. \ref{fig2} is denoted by dashed lines. The inset shows the probability density functions (pdf) of $\langle r_n\rangle$ for $\xi/\pi$ $=$ $1/2$ and $N$ $=$ $1001$ as an example, which presents a level repulsion without disorder ($\bar r$ $\approx$ $0.97$), in contrast to the case with $\bar w$ $=$ $0.5$ ($\bar r$ $\approx$ $0.4$), in the right and left narrowly distributed histograms respectively.}\label{fig5}
\end{figure}

Finally, we study the diagnostic measure of level statistics \cite{Abanin2019}, from which the averaged gap ratios exhibit Gaussian orthogonal ensemble (GOE) or Poisson distributions respectively in the nondisordered or disordered cases from a tight-binding model of interacting fermions \cite{Oganesyan2007, Sierant2019}. The gap ratio $r_n$ is determined by the adjacent gaps of ascendant eigenspectrum, $\delta_n$ $=$ $E_{n+1}$ $-$ $E_n$, which leads to a dimensionless $r_n$ $\equiv$ ${\rm min}\{\delta_n, \delta_{n-1}\}/{\rm max}\{\delta_n, \delta_{n-1}\}$ \cite{Oganesyan2007}. For each disorder realization, we define $r_a$ $\equiv$ $\sum_{n=2}^{N-1}r_n/(N-2)$ and obtain the mean gap ratio $\bar r$ $=$ $\langle r_a\rangle$. This shows a level repulsion in $\bar r_{\rm GOE}$ $\approx$ $0.53$, in contrast to $\bar r_{\rm Poisson}$ $\approx$ $0.39$ with uncorrelated energy levels owing to strong disorder. The intrasample  variance $\langle v_I\rangle$ $\equiv$ $\langle r_n^2-r_a^2\rangle$ can also be evaluated, which presents the fluctuations of level repulsions \cite{Sierant2019}. We then extract the gap ratios from the real parts of the eigenvalues obtained in the coupling matrix of Eq. \ref{coupling} at $D$ $=$ $0$ in particular, otherwise the matrix becomes defective, and eigen-decomposition fails \cite{Jen2020_subradiance}. In Fig. \ref{fig5}, the level repulsions and fluctuations respectively move to a lower and higher value, corresponding to the phase transition to localization. A probability density function of $\langle r_n\rangle$ for a large $N$ further presents a narrow distribution and clearly distinguishes two separate regimes of nondisordered and disordered systems, albeit neither GOE or Poisson statistics can apply in our system. This shows that the chirally coupled quantum emitters present a distinct long-range spectral correlation \cite{Sierant2019}, and the gap ratios and fluctuations obtained here find no similarity in other many-body spin models.  

{\it Discussions.} To realize disorder-assisted excitation localization in a chirally coupled atomic array, one potential platform is using an optical lattice near a nanofiber \cite{Corzo2019} with atoms loaded from a magneto-optical trap \cite{Gauroud2016, Solano2017_2} and controlling their coupling directionality by external magnetic fields \cite{Mitsch2014}. A successful demonstration of our results, however, will be limited by the system's nonradiative losses $\gamma_{nr}$. We use $\beta$ $\equiv$ $(\gamma_L+\gamma_R)/(\gamma_L+\gamma_R+\gamma_{nr})$ \cite{Arcari2014, Tiecke2014, Lodahl2017} to characterize the amount of the guided modes over the all including $\gamma_{nr}$ and provide a measure of system's performance. Considering a time window of at least $\gamma t$ $\sim$ $250$ to observe the phase transitions in Fig. \ref{fig1}, $\gamma_{nr}$ needs to be less than $\gamma/250$, which leads to $\beta$ $>$ $99.6\%$. This value looks stringent compared to the reported $\beta$ $>$ $90\%$ \cite{Mitsch2014}, but it can be improved with an external cavity \cite{Tiecke2014} or by implementing quantum dots on an optical fiber \cite{Yala2014} to achieve the strong coupling regime. Other potential platforms can be quantum dots in a waveguide  \cite{Arcari2014, Sollner2015} or superconducting qubits \cite{Roushan2017, Xu2018, Wang2020}, where the former has surpassed $\beta$ $=$ $98\%$, close to our estimation of requirement, and the latter has advantages of controlling system Hamiltonians and state preparations. We note that the latter has demonstrated the signatures of many-body localizations in a Bose-Hubbard \cite{Roushan2017} and a spin-$1/2$ $XY$ models \cite{Xu2018}, which can make a step further to simulate the localization phenomena in open quantum system with chiral couplings \cite{Hamann2018}.  

In conclusion, we study the long-time dynamics of a central excitation in chirally coupled quantum emitters under disordered potentials. We numerically obtain a dynamical phase boundary from excitation delocalization to localization, with dependences on disorder strengths, light-induced dipole-dipole interactions, and the directionality of couplings. We find an interaction-enabled delocalization and locate the phase regions for interaction-driven re-entrance of the localization phase. This dynamical phase corresponds to a decrease of the exponent of power-law decaying von Neumann entropy and manifests a reduction of gap ratio along with an increase in its fluctuations. The investigation of an excitation diffusion in disordered chirally coupled quantum emitters presents rich opportunities in studying nonequilibrium dynamics and restoration of quantum information, and paves the way toward a realization of many-body localization in open quantum systems.

{\it Acknowledgments.} We acknowledge support from the Ministry of Science and Technology (MOST), Taiwan, under the Grant No. MOST-106-2112-M-001-005-MY3 and insightful discussions with Y.-C. Chen, G.-D. Lin, and M.-S. Chang.


\end{document}